\documentclass[apj]{emulateapj}

\begin{document}
\title{Detection of VHE radiation from the BL Lac PG 1553+113
with the MAGIC telescope}
\author{
 J.~Albert\altaffilmark{a}, 
 E.~Aliu\altaffilmark{b}, 
 H.~Anderhub\altaffilmark{c}, 
 P.~Antoranz\altaffilmark{d}, 
 A.~Armada\altaffilmark{b}, 
 C.~Baixeras\altaffilmark{e}, 
 J.~A.~Barrio\altaffilmark{d},
 H.~Bartko\altaffilmark{g}, 
 D.~Bastieri\altaffilmark{h}, 
 J.~Becker\altaffilmark{f}
% ...
}
 \altaffiltext{a} {Universit\"at W\"urzburg, D-97074 W\"urzburg, Germany}
 \altaffiltext{b} {Institut de F\'\i sica d'Altes Energies, Edifici Cn., E-08193 Bellaterra (Barcelona), Spain}
 \altaffiltext{c} {ETH Zurich, CH-8093 Switzerland}
 \altaffiltext{d} {Universidad Complutense, E-28040 Madrid, Spain}
 \altaffiltext{e} {Universitat Aut\`onoma de Barcelona, E-08193 Bellaterra, Spain}
 \altaffiltext{f} {Universit\"at Dortmund, D-44227 Dortmund, Germany}
 \altaffiltext{g} {Max-Planck-Institut f\"ur Physik, D-80805 M\"unchen, Germany}
 \altaffiltext{h} {Universit\`a di Padova and INFN, I-35131 Padova, Italy} 
% ...

\begin{abstract}
In 2005 and 2006, the MAGIC telescope has observed very high energy gamma-ray emission
from the distant BL~Lac object PG 1553+113. The overall significance
of the signal is $8.8~ \sigma$ for 18.8~h observation time. The light
curve shows no significant flux variations on a daily time-scale, the
flux level during 2005 was, however, significantly higher compared to
2006. The differential energy spectrum between $\sim 90$~GeV and
500~GeV is well described by a power law with photon index $\Gamma = 4.2
\pm 0.3$. The combined 2005 and 2006 energy spectrum provides an
upper limit of $z=0.74$ on the redshift of the object.
\end{abstract}


\section{Introduction}

\subsection{The BL~Lac object PG 1553+113}
The Active Galactic Nucleus (AGN) PG 1553+113 was first reported in
the Palomar-Green catalogue of UV bright objects \citep{green}. It was
the only new BL~Lac object found in the survey and the first BL
Lac object found in an optical survey.  Its spectrum is, typical for
BL~Lac objects, featureless \citep{miller83} and the optical
variability strong ($m_p = 13.2-15.0$; \cite{miller88}). The spectral
characteristics are close to those of X-ray selected BL~Lacs
\citep{falomo90} and it is classified in the literature as
intermediate BL~Lac \citep{laurent99, nieppola06} or high-frequency
peaked BL~Lac \citep{giommi95}, as its synchrotron peak frequency
lies on the borderline of these two groups.

Despite several attempts, no emission or absorption lines have been
found in the spectrum of PG 1553+113 \citep{falomo90}. Thus only
indirect methods can be used to determine the redshift $z$
(e.g. \cite{sbarufatti05, sbarufatti06}). The host
galaxy was not resolved in Hubble Space Telescope (HST) images
\citep{urry00}, it is therefore safe to assume $z>0.25$. The
observation of very high energy (VHE, defined here as $E \gtrsim 100$
GeV) $\gamma$-ray emission, on the other hand, may permit to set an
upper limit on $z$. The $\gamma$-ray absorption in the Extragalactic
Background Light (EBL) by means of $e^+\ e^-$ pair production
\citep{stecker92, hessebl} can significantly affect the shape of the
observed energy spectrum depending on the source redshift. Based on
present-day EBL models and the observed $\gamma$-ray spectrum, one can
derive the intrinsic spectrum as a function of $z$. Physical
constraints on e.g. the slope of the intrinsic spectrum may then
permit to set upper limits on the possible redshift \citep{hess1553}.

PG 1553+113 belongs to a catalog of X-ray bright objects
\citep{donato05} and, based on its Spectral Energy Distribution (SED)
properties, was one of the most promising candidates from a list of
VHE $\gamma$-ray emitting candidates proposed by \cite{costamante02}.
So far, upper limits on the $\gamma$-ray emission have been reported
by the Whipple collaboration (19\% Crab flux above 390 GeV,
\cite{deperez03}) and Milagro \citep{williams04}. Recently the
H.E.S.S. collaboration has presented evidence for a $\gamma$-ray
signal at the $4\sigma$ level (up to $5.3\ \sigma$ using a low energy
threshold analysis) above 200 GeV corresponding to about 2\% of the
Crab flux \citep{hess1553}. The energy spectrum was found to have a
steep slope with $\Gamma = 4.0\pm0.6_{stat}$ and an upper limit on
the redshift of $z<0.74$ was derived.

\subsection{The MAGIC telescope}

The MAGIC telescope is located on the Canary Island of La Palma 
($28.75^{\circ}$~N, $17.86^{\circ}$~W, at 2225~m asl.). The telescope
comprises a 17~m diameter tessellated, parabolic mirror with a total
surface of 234~m$^2$, a light-weight space-frame made from carbon
fiber-epoxy tubes, and a camera with 576 hemispherical photo-multiplier
tubes (PMT) with enhanced quantum efficiency ~\citep{lacquer}. The
field of view of the camera is 3.5$^\circ$ while the trigger
area covers about 2.0$^\circ$ in diameter. The fast PMT analog signals
are routed via optical fibers to the DAQ-system electronics in the
counting house 80~m away. The signals are digitized by dual range
300~MHz FADCs. MAGIC can explore $\gamma$-rays at energies down to
50~GeV (trigger threshold, depending on the zenith angle), critical
for the observation of medium redshift VHE sources with steeply
falling spectra like PG 1553+113. The MAGIC telescope parameters and
performance are described in more detail in \cite{magiccomm} and
\cite{magictech}.\\
Simultaneous with MAGIC, optical observations were performed with the
KVA telescope on La Palma, operated remotely from Tuorla
Observatory. The main instrument is a 60~cm (f/15) Cassegrain
telescope equipped with a CCD capable of polarimetric measurements. A
35~cm auxiliary telescope (f/11) is mounted on the same RA axis. This
telescope is used for BVRI CCD photometry.

\section{\label{sect:analysis}Observation and data analysis}

PG 1553+113 was observed with the MAGIC telescope for 8.9~h in April
and May 2005, i.e. at about the same time when also H.E.S.S. observed
the source, and for 19~h from January to April 2006. In addition to
the observations with MAGIC and the 35~cm photometric telescope
simultaneous data in X-rays were taken with the All-Sky-Monitor on
board the RXTE satellite. These data are provided on the web at
http://heasarc.gsfc.nasa.gov/xte\_weather/.\\
Data taken during non-optimal weather conditions or affected by
hardware problems were excluded from the analysis. Also, only data
taken at small zenith angles $ZA< 30^\circ$ (corresponding to a low
energy threshold which is suitable for a steep energy spectrum) were
retained although measurements went up to $53^\circ$. After these
selection cuts, 7.0~h and 11.8~h of good data remained for 2005 
and 2006, respectively. Given the mean $ZA$ of $\sim 22^{\circ}$
$\gamma$-ray events above $\sim 90~$GeV have been used for the physics
analysis.\\
In addition to the so-called on-data from PG 1553+113, off-data were
taken on a nearby sky position where no $\gamma$-ray source is
expected, but with comparable zenith angle distribution and night sky
background light conditions. The off-data are used to determine the
background content in the signal region of the on-data. This was done
by means of a second order polynomial fit (without linear term) to the
ALPHA distribution of the normalized off-data. The normalization was
done in the $ALPHA$ region between $30^\circ$ and $90^\circ$ where no
$\gamma$-ray events are expected. The ALPHA parameter describes the
orientation of a shower image in the camera with respect to the camera
center. Air showers which are aligned parallel to the telescope axis do
have ALPHA values close to zero.\
In total, 14.5~h of off-data (6.5~h from 2005 and 8.0~h from 2006)
have been used for the analysis. Since the two off-samples were in
good agreement we used the combined data to analyze the individual
on-data samples.

The data were analyzed using the standard MAGIC analysis programs
for calibration, image cleaning, cut optimization and energy
reconstruction \citep{bretz05,gaug05,wagner05}. The primary method for
discrimination between hadron- and $\gamma$-ray-induced showers is
based on the Random Forest (RF) method \citep{breiman,bock04} which
was trained on off-data and Monte Carlo (MC) generated $\gamma$-ray
events. The significance of any excess was calculated according to
Eq.~17 in \cite{li83} where the on to off ratio $\alpha$ was derived,
taking into account the smaller error from the off-data fit. In
addition to the cut optimization, the RF method was also used for the
energy estimation based on the image parameters of a statistically
independent MC $\gamma$-ray sample. The average energy resolution
obtained was 24\% RMS. All MC data used in this analysis were
generated using the CORSIKA version 6.019 \citep{knapp04,majumdar05}.

\section{\label{sect:results}Results}

\begin{figure} [htb]
\epsscale{0.8}
\plotone{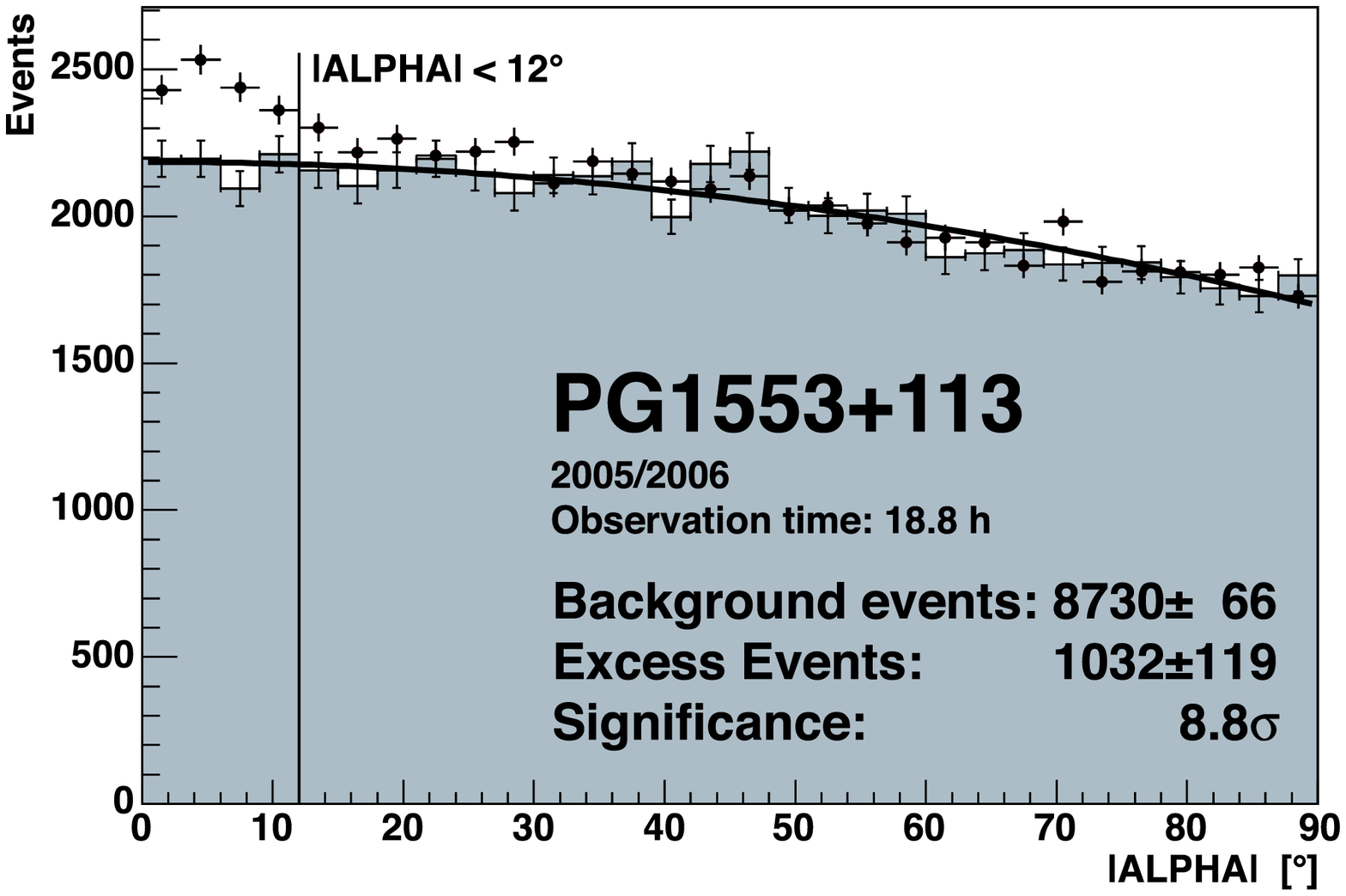}
\caption{ALPHA plot for the combined 2005 and 2006 PG
1553+113 data after cuts. The diagram also shows the distribution of
the (normalized) off-data and a second-order polynomial describing the
off-data.
\label{fig:alphaplot}}
\end{figure}

Combining the data from 2005 and 2006, a very clear signal is seen in
the image parameter ALPHA, as shown in
Fig.~\ref{fig:alphaplot}. Defining the signal region as $ALPHA <
12^\circ$ (containing about 90\% of the $\gamma$-ray events), an
excess of 1032 over 8730 background events yields a total significance
of $8.8~\sigma$. The individual results for the years 2005 and 2006
are listed separately in Table~\ref{tab:results}. In both years the
object has been clearly detected with a significance $> 6\sigma$.

\begin{figure}[htb]
\epsscale{1.0}
\plotone{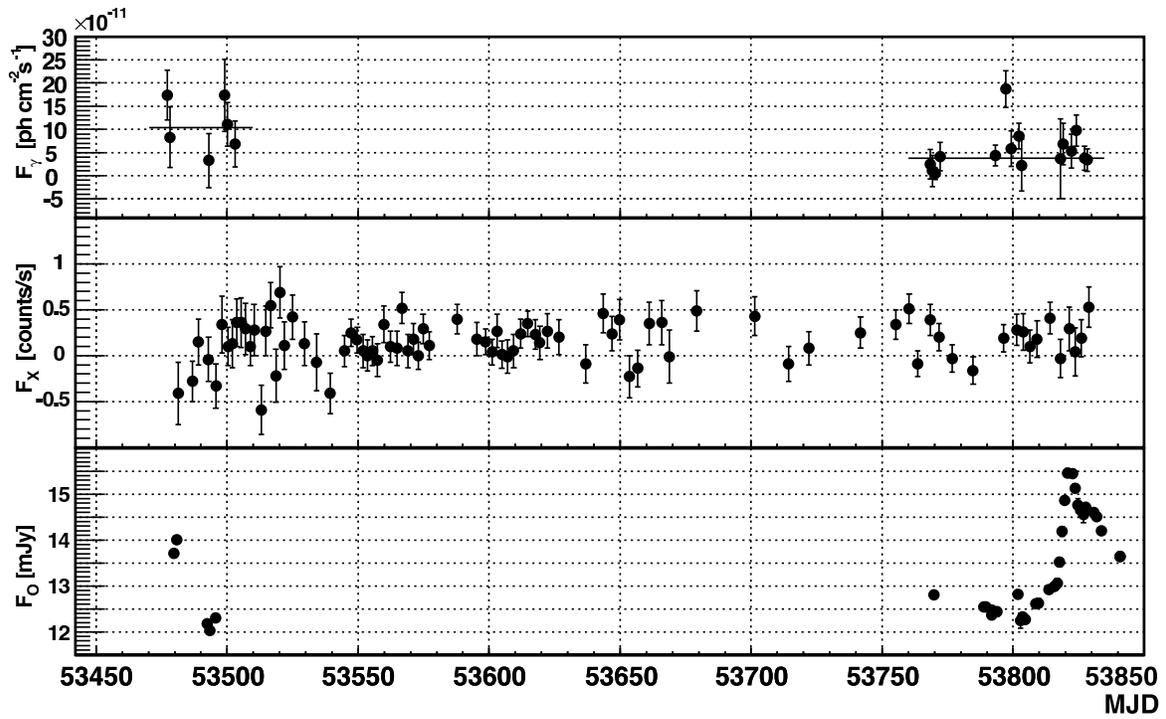}
\caption{VHE $\gamma$-ray ($120~\mathrm{GeV}$ -- $400~ \mathrm{GeV}$),
X-ray ($2~\mathrm{keV}$ -- $10~\mathrm{keV}$) and optical light curve
(R-Band) of PG 1553+113 in 2005 and 2006. The horizontal bars in
the top panel correspond to the average flux during 2005 and 2006,
respectively.
\label{fig:lightcuve}}
\end{figure}

The $\gamma$-ray, X-ray and optical light curve of PG 1553+113 are
shown in Fig.~\ref{fig:lightcuve}. While the optical data shows
significant short term variability on the 25\% level the X-ray data is
consistent with a constant emission, given the weighted mean of $0.15
\pm 0.03~\mathrm{counts/s}$. In $\gamma$-rays there is no evidence for
short term variability, but a significant change in the flux level
from 2005 to 2006 is found, given a systematic error of the analysis
on the flux level of about 30\%. The average integral flux between
$120~ \mathrm{GeV}$ and $400~ \mathrm{GeV}$ is given as $F = 10.0 \pm
0.23_{stat}$ and $F = 3.7 \pm 0.08_{stat}$ (with $F$ given in units
of $10^{-11}\ \mathrm{cm^{-2}\ s^{-1}}$) for 2005 and 2006,
respectively. On February 25th, prior to the optical flare, the
optical polarimetry of the source was measured with the KVA 60~cm
telescope. The degree of optical linear polarisation was $8.3 \pm
0.2\%$ and the polarisation position angle was $139.1^\circ \pm
0.4^\circ$. It should be noted, that since the host galaxy can not be
resolved for this object, the optical flux should correspond to the
emission from the AGN core.

\begin{figure}[htb]
\epsscale{1.0}
\plotone{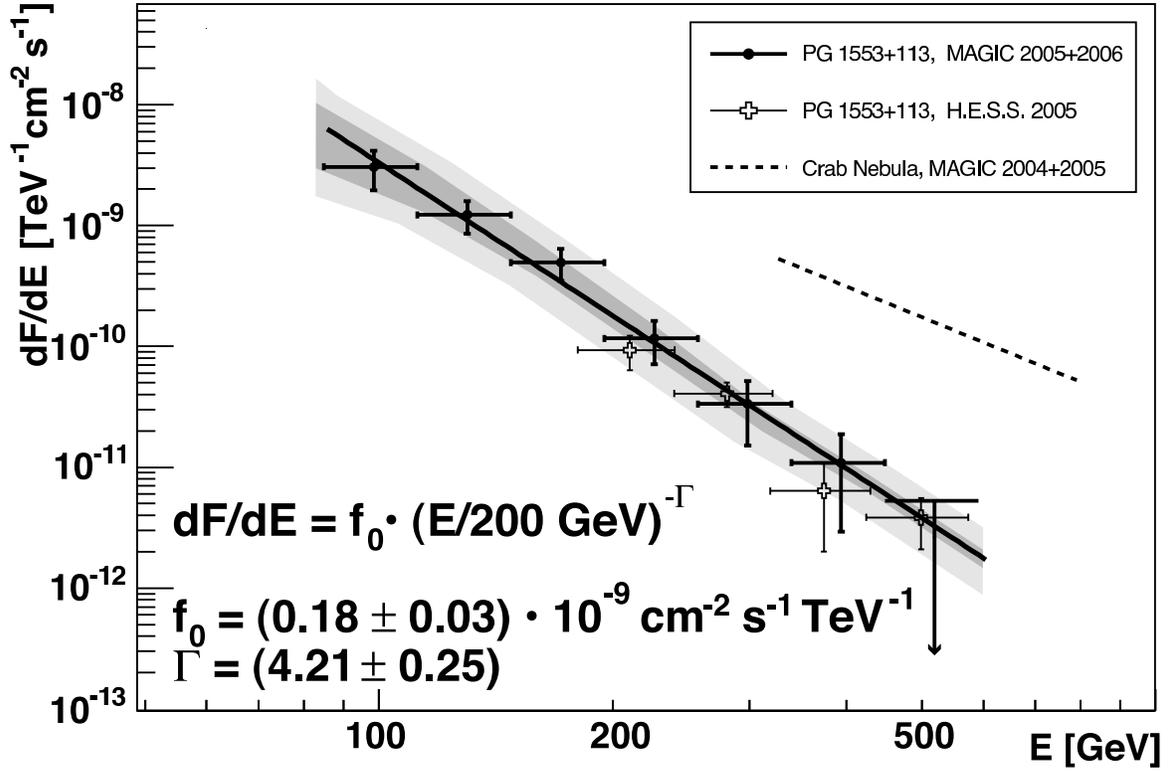}
\caption{Differential energy spectrum of PG 1553+113 as derived from
the combined 2005 and 2006 data. The MAGIC Crab energy spectrum and
the H.E.S.S. PG 1553+113 energy spectrum have been included for
comparison.
\label{fig:spectrum_1}}
\end{figure}

The combined 2005 and 2006 differential energy spectrum for PG
1553+113 is shown in Fig.~\ref{fig:spectrum_1}. The integral fluxes
above 200~GeV and the spectral slope coefficients for the different
samples are listed in Tab.~\ref{tab:results}. Effects on the spectrum
determination introduced by the limited energy resolution were
corrected by 'unfolding' according to \cite{mizobuchi05}. For
comparison, the MAGIC Crab energy spectrum and the H.E.S.S. PG
1553+113 energy spectrum \citep{hess1553} are also shown. The energy
spectrum is well described by a pure power law:

\begin{equation}
\frac{\mathrm{d}N}{\mathrm{d}E}=\left( 1.8 \pm 0.3_{stat}
\right) \cdot \left( \frac{E}{200\ \mathrm{GeV}} \right) ^{-4.2
\pm 0.3_{stat}}
\end{equation}

\noindent(in units of $10^{-10}\mathrm{cm^{-2}\ s^{-1} \ TeV^{-1}}$,
$\chi^2 / NDF = 1.5 / 4$). Compared to the Crab spectrum in the same
energy range ($\Gamma = 2.41\pm 0.05$; \cite{wagner05}) this spectrum
is significantly steeper. The spectral slopes of the individual years
are in good agreement although the flux level above $200~ \mathrm{GeV}$
is about a factor 3 larger in 2005 compared to 2006. This is also
shown in the light curve. The estimated systematic error on the
analysis (signal extraction, cut efficiencies etc.) is 25\% (dark
colored band in the figure) and 30\% on the energy scale (light
colored band).

\section{\label{sect:conclusion}Discussion}

The BL~Lac object PG 1553+113 has been detected at $8.8~ \sigma$ with
the MAGIC telescope in 18.8 hours of observation during 2005 and
2006. This confirms the tentative signal seen by H.E.S.S. at a higher
energy threshold with data taken at about the same time as MAGIC in
the 2005 period \citep{hess1553}. The source, therefore, can now be
considered as firmly detected.\\
The agreement between the measured H.E.S.S. and MAGIC energy spectra
of PG 1553+113 in 2005 is reasonably good. While the spectral slope is
consistent within errors, the absolute flux above 200~GeV in
2005 is by a factor 4 larger compared to H.E.S.S. This difference may
in part be explained by the systematic errors of both measurements but
also by variations in the flux level of the source (the observations
with H.E.S.S. were commenced after MAGIC). The observed energy
spectrum is steeper than that of any other known BL~Lac object. This
may be an indication of a large redshift ($z \gtrsim 0.3$), but can as
well be attributed to intrinsic absorption at the AGN or, more
naturally, to an inverse Compton peak position at lower energies.
The spectrum can, however, be used to derive an upper limit on the
source redshift from physical constrains on the intrinsic photon index
($\Gamma_{int} > 1.5$) as discussed in \cite{hess1553}. Using the
lower limit on the evolving EBL density from \cite{kneiske04} we
derived a $2 \sigma$ upper limit on the redshift of $z <
0.74$. The same value was reported by \cite{hess1553} where a
slightly different EBL model was used.

\begin{figure}[htb]
\epsscale{1.0}
\plotone{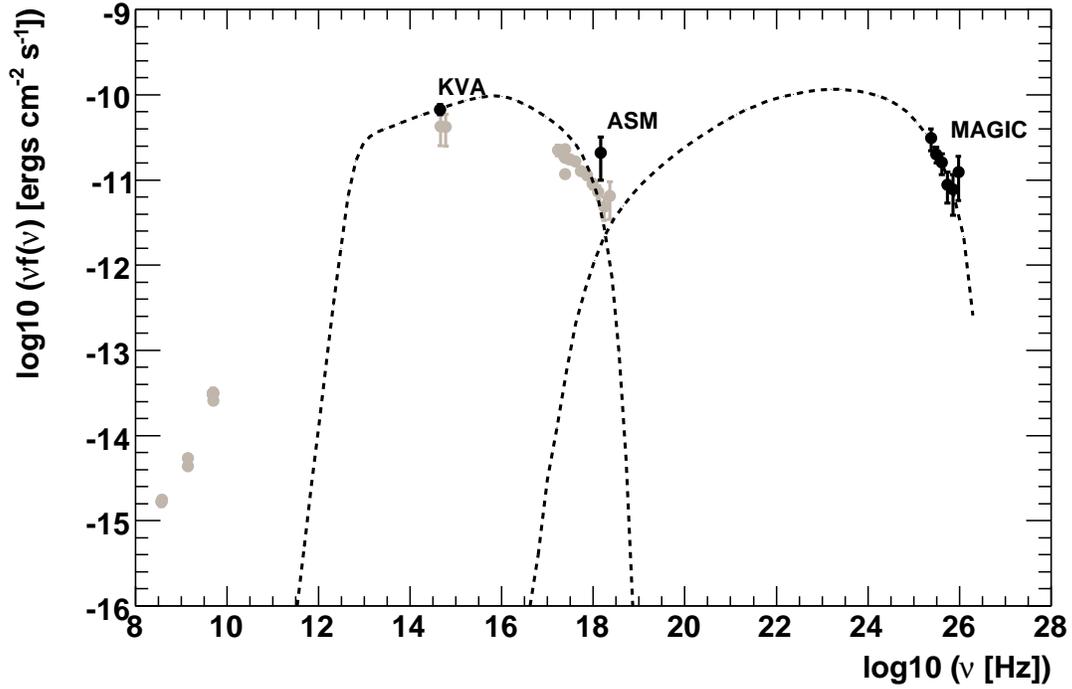}
\caption{Broad band SED of PG 1553+113. The solid lines are the result
of a SSC model fit to the black data points using the code provided by
\cite{krawczynski04} (see text). The gray points comprise
non-simultaneous radio, optical and X-ray data from \cite{giommi02}.}
\label{fig:spectrum_2}
\end{figure}

The Broad band SED of PG 1553+113 together with the results from a
model calculation are shown in Fig.~\ref{fig:spectrum_2}.The VHE data
points correspond to the intrinsic spectrum of PG 1553+113 as derived
for a redshift of $z = 0.3$. The black points at low energies
denote the average optical and X-ray flux taken at the same time as
the MAGIC observations. The gray hatched radio, optical and X-ray
non-simultaneous data were taken from \cite{giommi02}. The
solid line shows the result of a model fit to the simultaneously
recorded data (black points) using a homogeneous, one-zone Synchrotron
Self-Compton (SSC) model as provided by \cite{krawczynski04}. As can
be seen from Fig.~\ref{fig:spectrum_2}, the $\gamma$-ray, X-ray
and optical data are well described by the model. This is not the case
for the radio data where intrinsic absorption requires a much larger
emitting volume compared to X-rays and $\gamma$-rays. Except for a
somewhat smaller radius of the emitting region, identical model
parameters as in \cite{costamante02} have been used: Doppler factor
$D=21$, magnetic field strength $B=0.7\ \mathrm{G}$, radius of the
emitting region $R=1.16^{+0.62}_{-0.21} \cdot 10^{16}\ \mathrm{cm}$,
electron energy density $\rho_e = 0.11^{+0.18}_{-0.06} \ \mathrm{erg /
cm^3}$ slope of the electron distribution $\alpha_e=-2.6$ for $8.2 <
\log \left( E / \mathrm{eV} \right) < 9.8^{+0.2}_{-0.05}$ and
$\alpha_e=-3.6$ for $9.8^{+0.2}_{-0.05} < \log \left( E / \mathrm{eV}
\right) <10.6^{+1.6}_{-0.0}$. The limits on some of these parameters
indicate the change of the SED model parameters when varying the
assumed redshift from z=0.2 up to z=0.7 (parameters without limits
were kept constant for all fits). In the case of $z \ge 0.56$ the SED
model can not accurately describe the data and, based on the obtained
$\chi^2$ value, a redshift of 0.56 is excluded on the $4.5\sigma$
level. For a comparison of the model parameters with those from other
BL Lacs we refer to \cite{costamante02}.\\
PG 1553+113 was in a high state in the optical in both years
showing a strong flare at the end of March 2006. The high linear
polarization of the optical emission ($8.3 \pm 0.2 \%$) indicates that
a sizeable fraction of the optical flux is indeed synchrotron
radiation. In $\gamma$-rays only a significant change in the flux
level from 2005 to 2006 is found while there is no evidence for
variability in X-rays. As a result, a possible correlation between the
different energy bands can not be established. A possible connection
between the $\gamma$-ray detection and the optical high state can,
however, not be excluded. The optical flare without X-ray or
$\gamma$-ray counterpart may still be explained by
external-inverse-Compton (EIC) models which predict a time lag of the
X-rays and $\gamma$-rays with respect to the optical emission.

\section*{Acknowledgments}

We would like to thank the IAC for the excellent working conditions on 
the La Palma Observatory Roque de los Muchachos. We are grateful to the
ASM/RXTE team for their quick-look results.
The support of the German BMBF and MPG, the Italian INFN and the Spanish CICYT
is gratefully acknowledged. This work was also supported by ETH Research Grant 
TH-34/04-3 and by Polish Grant MNiI 1P03D01028.

\clearpage

\clearpage
\begin{deluxetable}{ccccccccc}
\tabletypesize{\scriptsize}
\tablecaption{Results from the PG 1553+113 analysis as derived for
2005 and 2006.\label{tab:results}}
\tablewidth{0pt}
\tablehead{
\colhead{Year} & \colhead{on time} & \colhead{N$_{on}$} &
\colhead{N$_{off}$} & \colhead{N$_{excess}$} & \colhead{on / off}
& \colhead{sigma} & \colhead{$F \left( E > 200 \mathrm{GeV}
\right)^a$} & \colhead{photon index}}
\startdata
2005 & 7.0~h & 3944 & $3501 \pm 26$ & $443 \pm 68$ & 0.20 & $6.7\
\sigma$ & $2.0 \pm 0.6_{stat} \pm 0.6_{sys}$ & $4.31 \pm 0.45$ \\
2006 & 11.8~h & 5815 & $5228 \pm  39$ & $588 \pm 86$ & 0.30 & $7.0\
\sigma$ & $0.6 \pm 0.2_{stat} \pm 0.2_{sys}$ & $3.95 \pm 0.23$ \\
2005+2006 & 18.8~h & 9761 & $8730 \pm 66$ & $1032 \pm 119$ & 0.49 &
$8.8\ \sigma$ & $1.0 \pm 0.4_{stat} \pm 0.3_{sys}$ & $4.21 \pm 0.25$ \\
\enddata
\tablenotetext{a}{integral flux in units of $10^{-11}\
\mathrm{cm^{-2}\ s^{-1}}$}
\end{deluxetable}


\begin{thebibliography}{10}
\bibitem [Aharonian et~al.(2006a)] {hessebl} Aharonian, F. {\it et
al.} 2006a, {\it Nature}, {\bf 440}, 1018
\bibitem [Aharonian et~al.(2006b)] {hess1553} Aharonian, F. {\it et
al.} 2006b {\it A\&A}, {\bf 448}, L19
\bibitem [Baixeras et~al.(2004)] {magiccomm} Baixeras, C. {\it et al.}
2004, {\it Nucl.Inst.Meth.}, {\bf A 518}, 188
\bibitem [Bock et~al.(2004)] {bock04} Bock, R.~K. {\it et al.} 2004,
{\it Nucl.Inst.Meth.}, {\bf A 516}, 511
\bibitem [Breiman (2001)] {breiman} Breiman, L. 2001, {\it Machine
Learning}, {\bf 45}, 5
\bibitem [Bretz et~al.(2005)] {bretz05} Bretz, T. {\it et al.} 2005,
AIP Conference Proceedings, {\bf 745}, 730
\bibitem [Costamante \& Ghisellini (2002)] {costamante02} Costamante L. \&
Ghisellini, G. 2002, {\it A\&A}, {\bf 384}, 56
\bibitem [Cortina et~al.(2005)] {magictech} Cortina, J. {\it et al.}
2005, Proc. 29th ICRC, Pune, 5-359, preprint (astro-ph/0508274)
\bibitem [de la Calle Perez et~al.(2003)] {deperez03} de la Calle
Perez, I. {\it et al.} 2003, {\it  ApJ}, {\bf 599}, 909
\bibitem [Donato et~al.(2005)] {donato05} Donato, D., Sambruna, R.~M.,
\& Gliozzi, M. 2005, {\it  A\&A}, {\bf 433}, 1163
\bibitem [Falomo \& Treves(1990)] {falomo90} Falomo, R. \& Treves,
A. 1990, {\it PASP}, {\bf 102}, 1120
\bibitem [Gaug et~al.(2005)] {gaug05} Gaug, M. {\it et al.} 2005,
Proc. 29th ICRC, Pune, 5-375, preprint (astro-ph/0508274)
\bibitem [Giommi et~al.(1995)] {giommi95} Giommi, P., Ansari, S.~G.,
\& Micol, A. 1995, {\it A\&AS}, {\bf 109}, 267
\bibitem [Giommi et~al.(2002)] {giommi02} Giommi, P., Massaro, E., \&
Palumbo, G. 2002, Proc. Blazar Astrophysics with BeppoSAX and other
Observatories, 2002, 63-100
\bibitem [Green et~al.(1986)] {green} Green, R.~F., Schmidt, M., \&
Liebert, J. 1986, {\it ApJS}, {\bf 61}, 305
\bibitem [Knapp \& Heck(2004)] {knapp04} Knapp, J. \& Heck,
D. 2004, EAS Simulation with CORSIKA: A Users manual
\bibitem [Kneiske et~al.(2004)] {kneiske04} Kneiske, T.~M., Bretz, T.,
Mannheim, K., \& Hartmann, D.~H. 2004, {\it A\&A}, {\bf 413}, 807
\bibitem [Krawczynski et~al.(2004)] {krawczynski04} Krawczynski, H.
{\it et al.} 2004, {\it ApJ}, {\bf 601}, 151
\bibitem [Laurent-Muehleisen et~al.(1999)] {laurent99}
Laurent-Muehleisen, S.~A., Kollgaard, R.~I., Feigelson, E.~D.,
Brinkmann, W., \& Siebert, J. 1999, {\it ApJ}, {\bf 525}, 127
\bibitem [Li \& Ma(1983)] {li83} Li, T. \& Ma, Y. 1983, {\it ApJ} {\bf
272}, 317
\bibitem [Majumdar et~al.(2005)] {majumdar05} Majumdar, P. {\it et
al.} 2005, Proc. 29th ICRC, Pune, 5-203, preprint (astro-ph/0508274)
\bibitem [Miller et~al.(1988)] {miller88} Miller, H.~R., Carini,
M.~T., Gaston, B.~J., \& Hutter, D.~J. 1988, Proc. of the IUE
Symposium (Greenbelt), 303
\bibitem [Miller \& Green(1983)] {miller83} Miller, H.~R. \& Green,
R.~F. 1983, {\it BAAS}, {\bf 15}, 957
\bibitem [Mizobuchi et~al.(2005)] {mizobuchi05} Mizobuchi, S. {\it et
al.} 2005, Proc. 29th ICRC, Pune, 5-323, preprint (astro-ph/0508274)
\bibitem [Nieppola et~al.(2006)] {nieppola06} Nieppola, E.,
Tornikoski, M., \& Valtaoja, E. 2006, {\it A\&A}, {\bf 445}, 441
\bibitem [Paneque et~al.(2004)] {lacquer} Paneque, D., Gebauer, H.~J.,
Lorenz, E., \& Mirzoyan, R. 2004, {\it Nucl.Inst.Meth.}, {\bf A 518} 619
\bibitem [Primack et~al.(2005)] {primack05} Primack, J., Bullock,
J.~S., \& Somerville, R.~S. 2005, AIP Conference Proceedings, {\bf 745}, 23
\bibitem [Sbarufatti et~al.(2005)] {sbarufatti05} Sbarufatti, B.,
Treves, A., \& Falomo, R. 2005, {\it  ApJ}, {\bf 635}, 173
\bibitem [Sbarufatti et~al.(2006)] {sbarufatti06} Sbarufatti, B.,
Treves, A., Falomo, R., Heidt, J., Kotilainen, J., \& Scarpa, R. 2006,
AJ, 132, 1
\bibitem [Stecker et~al.(1992)] {stecker92} Stecker, F.~W., de Jager,
O.~C., \& Salamon, M.~H. 1992, {\it ApJ} {\bf 390}, 49
\bibitem [Urry et~al.(2000)] {urry00} Urry, C.~M., Scarpa, R.,
O'Dowd, M., Falomo, R., Pesce, J.~E. \& Treves, A. 2000, {\it ApJ},
{\bf 532}, 816
\bibitem [Wagner et~al.(2005)] {wagner05} Wagner, R.~M. {\it et al.}
2005, Proc. 29th ICRC, Pune, 4-163, preprint (astro-ph/0508244)
\bibitem [Williams (2004)] {williams04} Williams, D. 2004, AIP
Conference Proceedings, {\bf 745}, 499
\end{thebibliography}
\end{document}